# Leveraging Generative Adversarial Networks for Addressing Data Imbalance in Financial Market Supervision

Mohan Jiang
New York University
New York, USA

Yaxin Liang
University of Southern California
Los Angeles, USA

Siyuan Han
Columbia University
New York, USA

Kunyuan Ma
New York University
New York, USA

Yuan Chen
Rice University
Houston, USA

Zhen Xu*
Independent Researcher
Shanghai, China

*Abstract*—This study explores the application of generative adversarial networks in financial market supervision, especially for solving the problem of data imbalance to improve the accuracy of risk prediction. Since financial market data are often imbalanced, especially high-risk events such as market manipulation and systemic risk occur less frequently, traditional models have difficulty effectively identifying these minority events. This study proposes to generate synthetic data with similar characteristics to these minority events through GAN to balance the dataset, thereby improving the prediction performance of the model in financial supervision. Experimental results show that compared with traditional oversampling and undersampling methods, the data generated by GAN has significant advantages in dealing with imbalance problems and improving the prediction accuracy of the model. This method has broad application potential in financial regulatory agencies such as the U.S. Securities and Exchange Commission (SEC), the Financial Industry Regulatory Authority (FINRA), the Federal Deposit Insurance Corporation (FDIC), and the Federal Reserve.

## I. INTRODUCTION

In the financial sector, data imbalance is a prevalent issue, particularly relevant in the context of regulatory monitoring by institutions like the SEC, FINRA, FDIC, and the Federal Reserve. These organizations are responsible for overseeing the stability and integrity of financial markets, ensuring that systemic risks are identified and mitigated. However, financial market data often exhibit significant imbalances, whether in the detection of fraudulent activities, market manipulation, or the identification of systemic risks [1]. For instance, events like significant market downturns or financial fraud occur infrequently but can have devastating impacts on market stability. This data imbalance poses challenges for traditional risk monitoring models, as they struggle to effectively predict these rare but critical events. The introduction of Generative Adversarial Networks (GAN) provides an innovative approach to addressing this imbalance by generating synthetic data that better represent rare events, thereby improving the performance of risk prediction models used by these regulatory bodies [2].

GANs have been extensively applied across various domains, demonstrating significant potential in advancing different research areas. In the field of UI design [3], GANs facilitate the automated generation of aesthetically pleasing and functional user interfaces by learning design patterns and preferences [4]. In knowledge reasoning [5], GANs contribute to the development of models capable of generating structured insights, thereby improving decision-making processes and enhancing inference capabilities [6]. Additionally, in computer vision [7], GANs are employed for numerous tasks, including image synthesis, style transfer, and the enhancement of image quality, contributing to the state-of-the-art in image processing and analysis [8]. These applications illustrate the versatility and efficacy of GANs in tackling complex challenges across multiple disciplines. In financial market regulation, GANs can create synthetic data representing rare but high-impact market events, such as systemic collapses or periods of extreme volatility. This approach is valuable for regulatory bodies like the Federal Reserve or SEC, as it improves the accuracy of models monitoring systemic risks and market integrity. The generation of such synthetic data helps regulators develop more comprehensive datasets, enhancing their ability to predict and respond to potential market threats.

Financial market risk prediction is vital for regulatory bodies to identify systemic risks, detect market manipulation, and forecast volatility. Traditional models often fail to capture low-frequency, high-impact events due to data imbalance, limiting early warnings of market crashes or fraud. GANs address this challenge by generating synthetic data that enhances model sensitivity to these minority events [9], improving regulators' ability to issue timely warnings. GAN-generated data simulates scenarios such as market crashes or fraud, providing regulatory agencies with valuable insights to stabilize financial markets [10].

GANs also enhance the robustness of risk monitoring. Financial markets, influenced by variables like policy changes, macroeconomic shifts, and global events, present highly volatile and complex data [11]. Traditional models often become unstable or overfit under these conditions [12]. GAN-generated data creates diverse datasets, enabling consistent model performance across different market environments. This helps institutions like the SEC and FDIC maintain accurate monitoring systems and better anticipate financial risks in volatile markets.

## II. RELATED WORK

Generative Adversarial Networks (GANs) have been widely recognized for their ability to tackle data imbalance and improve the performance of predictive models. This study builds on prior advancements in deep learning methodologies that address similar challenges, incorporating relevant insights to enhance risk prediction in financial market regulation.

Yao et al. [13] demonstrated the advantages of combining deep learning architectures for improving prediction accuracy, highlighting the importance of leveraging complementary model capabilities. Their work informs the approach taken in this study, where GANs are employed to augment datasets and improve model robustness.

Similarly, Wu et al. [14] emphasized the significance of optimizing feature interaction to handle complex data patterns, a concept closely related to generating synthetic data that captures rare yet impactful events in financial markets. Liu et al. [15] presented techniques for refining feature representations through advanced data processing mechanisms, such as self-attention. This aligns with the use of GANs in this paper to create high-quality synthetic datasets, which are critical for mitigating the impact of data imbalance. Additionally, Li et al. [16] explored adaptive learning methods for handling dynamic data conditions, providing complementary insights into improving model generalization and predictive stability.

The transformation of complex data into actionable formats has also been highlighted as a key area of focus. Yan et al. [17] proposed methodologies for extracting interpretable insights from multidimensional data, underscoring the importance of enhancing data usability for predictive tasks. These insights are directly applicable to this study's use of GANs to generate realistic synthetic data for rare financial events. Wei et al. [18] contributed methods for enriching data diversity and enhancing feature extraction [19], demonstrating the value of robust and representative datasets in improving model performance. The current study extends these findings by employing GANs to simulate rare market conditions, enabling more effective detection of systemic risks and market manipulation. Du et al. [20] explored strategies for handling complex data distributions, a challenge that resonates with this paper's focus on addressing data imbalance in financial datasets. Their emphasis on improving model sensitivity to underrepresented patterns aligns with the goals of synthetic data generation in financial market supervision.

Collectively, these works provide a foundation for applying GANs to generate synthetic data that addresses the limitations of traditional risk prediction models. By leveraging these contributions, this study enhances the ability of financial regulatory institutions to identify and respond to rare, high-impact events, ultimately improving the stability and integrity of financial markets.

## III. METHOD

In the process of using generative adversarial networks (GAN) to enhance financial data for regulatory market monitoring and improve risk prediction, the method design must be rooted in both mathematical reasoning and deep learning architecture. Regulatory bodies like the SEC, FINRA, FDIC, and the Federal Reserve require advanced technologies that can generate synthetic data to help address imbalances in financial datasets, particularly those imbalances associated with rare but impactful market events, such as systemic crashes or market manipulation. By generating such data, GAN can significantly enhance the capability of these institutions' monitoring systems, ensuring that they can better detect and respond to potential financial threats. The architecture diagram of the generative adversarial network is shown in Figure 1.

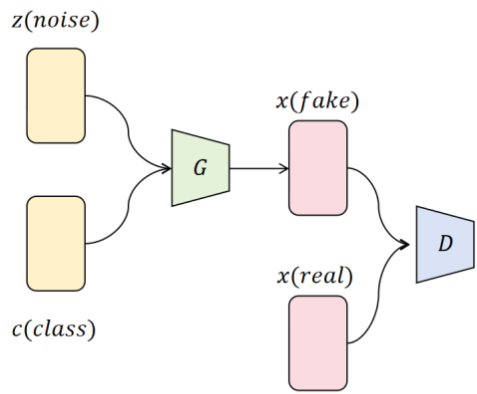

Figure 1 Generate adversarial network architecture diagram

GAN consists of two key components: the generator and the discriminator. The generator's objective is to produce synthetic financial data that mirrors the distribution of real-world market data, particularly focusing on rare events that are of high importance for regulatory oversight. The discriminator's role is to assess whether the input data is real or synthesized. During the training process, these two networks engage in an adversarial game where the generator becomes progressively better at producing realistic data, while the discriminator improves its ability to distinguish between genuine and synthetic data. This iterative process leads to the generation of synthetic data that closely resembles real financial data, particularly in areas of regulatory concern, such as market anomalies or systemic risks. Mathematically, this can be framed as a minimization-maximization problem, where the generator aims to minimize the difference between the synthetic and real data distributions, while the discriminator attempts to maximize its accuracy in detecting synthetic data. This approach provides regulatory bodies with more robust and realistic data for financial risk monitoring.

$$\min \max V(D,G) = E_{x \sim p_{data(x)}}[\log D(x)] + E_{z \sim p_{z(z)}}[\log(1 - D(G(z)))]$$

Among them, $G(z)$ represents the data generated by the generator based on the noise $z$, $D(x)$ represents the probability estimate of the discriminator on whether the input data $x$ is real data. $p_{data(x)}$ is the distribution of real data, and $p_{z(z)}$ is the noise distribution of the generator input, which is usually uniform distribution or normal distribution.

The goal of the generator is to make $D(G(z))$ close to 1, which means that the discriminator cannot distinguish between generated data and real data; while the goal of the discriminator is to maximize the discrimination effect, making $D(x)$ as close to 1 as possible and $D(G(z))$ as close to 0 as possible.

In financial market regulation, data often displays significant imbalances, particularly when monitoring high-risk events such as market crashes or systemic failures. These events are considerably rarer compared to normal market operations. Regulatory institutions, including the SEC, FINRA, FDIC, and the Federal Reserve, face substantial challenges when applying traditional machine learning models to such datasets. These models typically favor the majority class (normal market activities), making it difficult to accurately predict minority class events like market manipulation, fraud, or systemic crashes. To address this limitation, Generative Adversarial Networks (GANs) are increasingly utilized to create synthetic data that mirrors the characteristics of these rare minority class events. By augmenting the dataset with such synthetic samples, GANs help balance the data distribution, enabling regulatory bodies to enhance the detection and prediction of rare but critical financial risks. This advancement significantly improves the effectiveness of market oversight and risk assessment models, providing a robust mechanism for mitigating systemic vulnerabilities in financial markets.

Assuming that the distribution of minority class data samples is $p_m(x)$, we hope that the data $G(z)$ generated by generator $G$ satisfies the following:

$$p_g(x) = p_m(z)$$

That is, the generated data is as close to the distribution of the minority class as possible. Through such a generation process, more minority class samples can be provided to the model, alleviating the prediction bias problem caused by data imbalance.

During the training process, the optimization process of the generator and the discriminator can be performed through the stochastic gradient descent optimization algorithm. The loss function of the generator is:

$$L_G = -E_{z \sim p_z(z)}[\log D(G(z))]$$

The loss function of the discriminator is

$$L_D = -E_{x \sim pdata(x)}[\log D(x)] - E_{z \sim p(z)}[\log(1 - D(G(z)))]$$

By alternatively optimizing the loss functions of the generator and discriminator, we can gradually refine the generated data to closely resemble the distribution of the minority class in the real-world data. This approach effectively addresses the issue of financial data imbalance by employing generative adversarial networks for data augmentation. This technique enables the model to allocate its focus more equitably between majority and minority class events, thereby significantly enhancing the performance of financial risk prediction models.

## IV. EXPERIMENT

### A. Datasets

The dataset employed in this research pertains to financial market data pertinent to regulatory oversight. It encompasses records from market transactions, institutional reports, and financial filings monitored by regulatory bodies such as the Securities and Exchange Commission (SEC), Financial Industry Regulatory Authority (FINRA), Federal Deposit Insurance Corporation (FDIC), and the Federal Reserve. Specifically, the dataset draws from various sources, including the SEC EDGAR Database, which houses public company filings, including quarterly and annual financial statements furnished by the U.S. Securities and Exchange Commission (SEC); FINRA TRACE (Trade Reporting and Compliance Engine), which provides comprehensive transaction data on bonds, maintained by FINRA; and Federal Reserve Economic Data (FRED), administered by the Federal Reserve Bank of St. Louis, which offers a comprehensive repository of financial and economic data invaluable for financial market research and regulatory analysis. The primary objective of this dataset is to forecast market irregularities, encompassing instances of fraud, systemic risk events, or market manipulation. It comprises thousands of samples, each representing a distinct financial entity or transaction, and incorporates over 20 features that capture pertinent variables, such as transaction volume, market volatility, company financial health, regulatory filings, and other critical financial indicators. The dataset, while comprehensive, reflects a significant imbalance, with a far smaller number of high-risk events (e.g., fraud or market crashes) compared to normal market activities. This imbalance presents a challenge to traditional machine learning models, which tend to underperform in predicting these rare but high-impact events. To address this, we use Generative Adversarial Networks (GAN) to generate synthetic data representing minority events, such as market manipulation or fraud, thereby enriching the dataset. The diverse and complex nature of this data makes it highly suitable for developing and testing risk prediction models aimed at financial market regulation.

*B. Experimental Results*

To validate the effectiveness of the proposed method, this study utilized two machine learning algorithms—Random Forest and XGBoost—and two deep learning algorithms—Multilayer Perceptron (MLP) and Long Short-Term Memory (LSTM). Furthermore, the performance of the proposed model was evaluated by comparing it against traditional sampling techniques, including oversampling and undersampling methods.

Table 1 Experimental Results

| Model | Acc | F1 score |
|---|---|---|
| Random Forest | 69.3 | 68.7 |
| Random Forest+ours | 70.1 | 69.5 |
| Xgboost | 71.0 | 70.5 |
| Xgboost+ours | 72.1 | 71.9 |
| MLP | 73.2 | 73.1 |
| MLP+ours | 75.4 | 75.6 |
| LSTM | 78.6 | 77.9 |
| LSTM+ours | 80.5 | 79.8 |

From the experimental results, by introducing synthetic data generated by generative adversarial networks (GAN) in different models, there is a significant performance improvement, especially in the prediction of minority classes (such as defaulting customers) with data imbalance. First, in all the baseline results using traditional models, although the accuracy (Acc) and F1 score of the model are relatively high, due to the small number of defaulting customers in the original data set, the model tends to be more accurate for the majority class (non-defaulting customers), while the prediction performance for the minority class is weak. This can be seen from the baseline results of Random Forest and Xgboost. Although their accuracy rates are 69.3% and 71.0% respectively, their F1 scores (68.7% and 70.5%) are slightly insufficient, indicating that their performance in minority class prediction still has room for improvement.

After introducing synthetic data generated by GAN, the performance of all models has improved to varying degrees.

The accuracy of Random Forest+ours increased to 70.1%, and the F1 score also increased to 69.5%, showing that the enhancement of minority class samples has improved the model's ability to identify defaulting customers. Similarly, the performance of Xgboost+ours has also improved similarly, with accuracy increasing from 71.0% to 72.1% and F1 score increasing from 70.5% to 71.9%. These improvements show that by generating more minority class data through GAN, the model can better balance the majority class and minority class during training, thus performing better in dealing with imbalanced problems, especially for data-sensitive models like Xgboost.

Further analysis of the performance of the MLP and LSTM models, especially after the introduction of GAN, their performance improvement is more significant. Although the baseline results of MLP have reached 73.2% accuracy and 73.1% F1 score, after using synthetic data generated by GAN, the accuracy of MLP+ours has increased to 75.4%, and the F1 score has increased to 75.6%, showing that adding data in minority class samples greatly improves the generalization ability of the model. LSTM performed the strongest in the experiment, with a baseline result of 78.6% accuracy and 77.9% F1 score, but after the introduction of GAN, the accuracy of LSTM+ours jumped to 80.5% and the F1 score also increased to 79.8%. This shows that for more complex deep learning models, synthetic data generated by GAN provides more significant improvements in dealing with imbalanced problems, thereby greatly improving the performance of the model in predicting financial risks and financial supervision.

In summary, the experimental results show that the accuracy and F1 scores of all models have been improved by synthetic data generated by GAN, especially the prediction performance of minority class samples has been significantly enhanced. Traditional models (such as Random Forest and Xgboost) are limited by the lack of minority class samples when dealing with imbalanced data, but through data enhancement, these models can better capture the characteristics of minority classes, and the prediction performance is also improved. For complex models such as MLP and LSTM, the introduction of GAN makes the model perform well in dealing with financial risk prediction, showing the strong potential of GAN in imbalanced data problems.

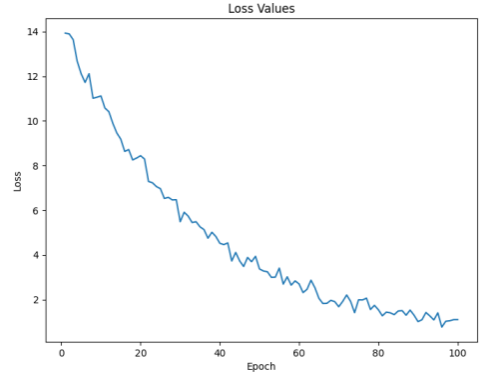

Figure 2 Loss Function Decrease Chart

To further illustrate our experimental process, we also give a graph of the loss function drop during the training process, as shown in Figure 2. In addition, we also analyzed the different results obtained by different methods to solve data imbalance. As shown in Table 2.

Table 2 Different methods to solve data imbalance

| Method | Acc | F1 score |
|---|---|---|
| Original(LSTM) | 75.5 | 74.9 |
| Under-sampling(LSTM) | 76.3 | 77.7 |
| Over-sampling(LSTM) | 78.7 | 78.5 |
| SMOTE(LSTM) | 79.0 | 78.5 |
| GAN(LSTM) | 80.5 | 79.8 |

From the experimental results presented in Table 2, it becomes evident that various data processing methods for imbalanced datasets have a substantial impact on the performance enhancement of the LSTM model. Initially, without any data enhancement or processing, the accuracy of the original LSTM model reached 75.5%, and the F1 score attained 74.9%. This outcome demonstrates that while the model effectively predicts the majority class samples in most instances, the performance is slightly compromised due to the relatively limited number of minority class samples in the dataset. Consequently, the slightly lower F1 score underscores the necessity for enhancing the accuracy and recall of the model when processing imbalanced data. This observation presents a research direction for further data enhancement. Data processing methods demonstrated their effectiveness in addressing data imbalance. Undersampling effectively balanced the dataset by reducing the majority class samples, resulting in an F1 score of 77.7% with a modest increase in accuracy to 76.3%. Oversampling and SMOTE exhibited superior performance, elevating accuracy to 78.7% and 79.0%, respectively, and both achieved an F1 score of 78.5%. Notably, the most optimal performance was achieved through the utilization of Generative Adversarial Networks (GAN) for data augmentation. GAN-generated synthetic data enhanced the quality and quantity of minority class samples, leading to an accuracy of 80.5% and an F1 score of 79.8%. This outcome signifies improved stability in handling imbalanced data.

## V. CONCLUSION

This paper studies the application of generative adversarial networks to deal with data imbalance in financial market supervision. Traditional financial market supervision models usually perform poorly in dealing with a few high-risk events, mainly because these events account for a very small proportion of the data set, causing the model to tend to overfit to routine events. The synthetic data generated by GAN effectively solves this problem. The generator of GAN can simulate a few events in the financial market and generate synthetic data with a distribution similar to that of real data, thereby expanding the range of data used to train the model. On the other hand, the discriminator ensures that the generated data is more realistic by continuously optimizing its discriminative ability. Through this adversarial mechanism, the generated data not only enhances the diversity of data but also significantly improves the prediction accuracy and stability of the model for a few events. Experimental results show that the data generated by GAN shows great improvement in a few events that are difficult to identify in traditional models, especially in the detection of systemic risk and market manipulation. Compared with traditional data augmentation methods, the data generated by GAN is closer to the real market environment and significantly improves the generalization ability of the model. Based on this, this study provides a feasible technical path for future financial market supervision, especially in a complex and changing market environment. By generating high-quality synthetic data, regulators can more effectively monitor and predict market risks. The widespread application of this technology will help enhance the stability of financial markets and reduce systemic risks.